\documentclass[aps,prx,twocolumn,showpacs,superscriptaddress,floatfix,nofootinbib]{revtex4-2}

\usepackage{graphicx,graphics}
\usepackage{dcolumn}
\usepackage{amsmath,amssymb,amsfonts}
\usepackage{latexsym,verbatim}
\usepackage{bm}
\usepackage{lipsum} 
\usepackage{dsfont}
\usepackage{cancel}
\usepackage{breqn}
\usepackage{bbm} 
\newcommand\bbone{\ensuremath{\mathbbm{1}}}
\makeatletter
\let\cat@comma@active\@empty
\makeatother

\usepackage[toc,page]{appendix}
\usepackage{color}
\usepackage{ulem}
\usepackage{tikz}
\usepackage[breaklinks=true,colorlinks,citecolor=blue,linkcolor=blue,urlcolor=blue]{hyperref}
\usepackage{tabularx}
\usepackage{array, multirow, bigdelim, makecell, booktabs} 
\graphicspath{fig}

\begin{document}
\title{Phenomenological Theory of Variational Quantum Ground-State Preparation}

\author{Nikita~Astrakhantsev}
\email[]{nikita.astrakhantsev@physik.uzh.ch}
\affiliation{Department of Physics, University of Zurich, Winterthurerstrasse 190, CH-8057 Z\"{u}rich, Switzerland}

\author{Guglielmo~Mazzola}
\affiliation{IBM Quantum, IBM Research --- Zurich, CH-8803 R\"{u}schlikon, Switzerland}
\affiliation{Institute for Computational Science, University of Zurich, Winterthurerstrasse 190, 8057 Zurich, Switzerland}

\author{Ivano~Tavernelli}
\affiliation{IBM Quantum, IBM Research --- Zurich, CH-8803 R\"{u}schlikon, Switzerland}

\author{Giuseppe~Carleo}
\affiliation{Institute of Physics, École Polytechnique Fédérale de Lausanne (EPFL), CH-1015 Lausanne, Switzerland}

\begin{abstract}
The variational approach is a cornerstone of computational physics, considering both conventional and quantum computing computational platforms.
The variational quantum eigensolver (VQE) algorithm aims to prepare the ground state of a Hamiltonian exploiting parametrized quantum circuits that may offer an advantage compared to classical trial states used, for instance, in quantum Monte Carlo or tensor network calculations.
While traditionally, the main focus has been on developing better trial circuits, we show that the algorithm's success crucially depends on other parameters such as the learning rate, the number $N_s$ of measurements to estimate the gradient components, and the Hamiltonian gap $\Delta$.
We first observe the existence of a finite $N_s$ value below which the optimization is impossible, and the energy variance resembles the behavior of the specific heat in second-order phase transitions.
Secondly, when $N_s$ is above such threshold level, and learning is possible, we develop a phenomenological model that relates the fidelity of the state preparation with the optimization hyperparameters as well as $\Delta$. More specifically, we observe that the computational resources scale as $1/\Delta^2$, and we propose a symmetry-enhanced simulation protocol that should be used if the gap closes.
We test our understanding on several instances of two-dimensional frustrated quantum magnets, which are believed to be the most promising candidates for near-term quantum advantage through variational quantum simulations.

\end{abstract}
\maketitle
\section{Introduction}
The task of preparing a quantum state on a qubit register is of fundamental importance in quantum computing. To this end, the Variational Quantum Eigensolver (VQE)~\cite{Peruzzo_2014,Moll_2018,cerezo2021variational} algorithm, of widespread use in several quantum application domains, from chemistry and physics~\cite{Kandala_2017,hempel2018quantum,Kokail2019}, machine learning~\cite{havlivcek2019supervised}, combinatorial optimization~\cite{farhi2014quantum}, and finance~\cite{Chakrabarti2021thresholdquantum} aims at variationally preparing the ground state of a Hamiltonian $\hat{H}$~\cite{https://doi.org/10.48550/arxiv.2111.05176,Cerezo_2021,Kandala_2017,Cao_2019,wecker2015progress,Chakrabarti2021thresholdquantum,cerezo2020variational,lubasch2020,mazzola2021sampling}. 
While the VQE algorithm has been implemented already on existing noisy devices, its importance will persist in the future fault-tolerant regime as it is also necessary for more advanced quantum algorithms such as eigenstate projection methods~\cite{abrams1999quantum,Reiher_2017}.

The two essential ingredients of the method are:
(i) a parametrized quantum circuit with variational parameters $\boldsymbol{\theta}$, which produces a wave function $|\psi(\boldsymbol \theta) \rangle$ expressing the ground state of the problem, and 
(ii) a \textit{learning} procedure aimed to optimizing the circuit variational parameters $\boldsymbol{\theta}$. This last step involves a feedback loop between the quantum and classical resources.

The VQE is among the most widely used quantum algorithms, and it has been adapted to many particular contexts and models. 
Significant effort has been spent so far in devising better variational forms, tailored to various application domains~\cite{Bartlettt_2007,wecker2015progress,adaptvqe,uccsd,cerezo2021variational,Kokail2019,Giulia_gauge2021,Seki_2020} or hardware architectures~\cite{Kandala_2017,Kokail2019}. Besides, optimization part of the algorithm has been studied~\cite{Cerezo_2021}, including works discussing the concept of the barren plateaus, manifested in gradients vanishing exponentially fast with the system size~\cite{cerezo2021cost,barren2018}.
Importantly, his phenomenon was reported only in the context of random circuit variational forms. Recent works have proven an absence of such plateaus in the quantum CNN architecture~\cite{2021_Pesah} and circuits aimed at the study of quantum magnets~\cite{https://doi.org/10.48550/arxiv.2108.08086,Liu_2019}.

However, significantly less attention has been paid to a realistic scenario of optimization with sizable shot noise, and to relation between the optimization efficiency and the problem's physical regime, including the dependency on the system gap $\Delta$~\cite{cerezo2021cost}.

This article aims to provide a phenomenological theory underlying the efficiency of the training (i) within several optimization algorithms, (ii) in various regimes of model parameters, and (iii) as a function of optimization hyperparameters such as the learning rate $\eta$ or the number of gradient-estimating samples $N_s$ per optimization step.

We adopt the best-case scenario of a noise-free hardware setting, i.\,e. we consider perfect gates and read-out, and study the stochastic gradient descent (SGD) and the stochastic reconfiguration (SR) approach~\cite{Sorella_1998}, as implemented on quantum computers in the Quantum Natural Gradient Descent (QNGD) approach~\cite{stokes_quantum_2020}. In such a hardware-free noise setting, the remaining obstacle for the state preparation is the inherent statistical quantum measurement noise in the gradient estimation ($\boldsymbol \nabla \langle \hat H \rangle$) induced by a finite budget $N_s$ of shots, i.\,e. circuit repetitions~\cite{wecker2015progress,Torlai_2020,cerezo2021variational}.

In this work, only consider gradient-based optimization. These methods are likely the only scalable approach to VQE as, at least in the classical setting, they represent the only practical strategy that provides stable optimization of a large number of parameters~\cite{nakano2020turborvb,Becca2017,https://doi.org/10.48550/arxiv.1911.02590}. Applications of other gradient-free methods such as SPSA, are restricted to the problems of quantum chemistry with a small number of variational parameters only~\cite{Cerezo_2021}.

In this scenario, we analyze how the performance of variational ground-state preparation is affected by the sample size $N_s$ per gradient component and the learning rate $\eta$ of a gradient descent step. First, we observe that the resulting fidelity shows a ``critical'' behavior as a function of a suitably-defined stochastic energy fluctuations measure $\epsilon \propto 1 / N_s$. Namely, we show that the fidelity vanishes when this measure is larger than a certain threshold value, $\epsilon > \epsilon^c$ and shows a rapid growth instead at $\epsilon < \epsilon^c$. {Further in the text, we refer to this sharp change in the algorithm performance as an {\it algorithmic phase transition} and, somewhat loosely, call this behavior {\it critical}. We emphasize that these algorithm performance regimes, as well as the transition between them and threshold (critical) $\epsilon_c$, stem primarily from the optimization hyperparameters such as $\eta,\,N_s$, and are not related to conventional phase transitions in a medium.} Notably, this critical behavior can be qualitatively reproduced for a given circuit with a simple distribution over the parameter space given by the Boltzmann distribution $\Pi(\boldsymbol \theta) \propto \exp \left(-E(\boldsymbol \theta) / T \right)$ with $T$ being the effective temperature of the system and $E(\boldsymbol \theta) = \langle \psi(\boldsymbol \theta)|\hat{H}| \psi(\boldsymbol \theta) \rangle$ being the energy expectation. Second, we address estimating the sample size required to reach a certain overlap with the ground state. For sufficiently large samples, we observe that the circuit infidelity behaves as $A \epsilon + \mathcal{I}_0$ with $\mathcal{I}_0$ representing the circuit's inability to exactly express the quantum state.

By considering numerical simulations of different two-dimensional frustrated spin-$1/2$ magnets, we observe that the prefactor $A$ has a universal behavior and, in the case of such systems, grows as $\sim 1 / \Delta^2$ with the spectral gap $\Delta$. Such dependence imposes a constraint on the minimum circuit shots number $N_s$ required to reach certain fidelity and on the class of quantum systems addressable with VQE. Based on this observation, we also discuss symmetry-based strategies to effectively increase $\Delta$, thus, in some cases, significantly reducing the required resources for the algorithm.

Before moving forward, let us comment on the choice of two-dimensional frustrated spin--$1/2$ models as the research focus. These models, to our belief, are among the most promising candidates for near-term quantum advantage for two reasons. (i) First, among the spin models, they represent the most challenging systems for classical calculations. This is in contrast to the one-dimensional or non-frustrated models,
in which the ground state can be efficiently computed.
(ii) Second, their qubit Hamiltonian remains local, unlike the typical quantum chemistry operators given by the sum of Pauli strings with arbitrary large support~\cite{Reiher_2017,Cao_2019}.
This enables quantum circuits containing only local gates, which is compatible with the current superconducting quantum platforms~\cite{48651}. Moreover, the gradient descent optimization in such problems requires measurement of a much smaller number of Hamiltonian terms', leading to
shorter runtimes compared to fermionic models.
Therefore, we believe the assessments in this manuscript are 
directly related to a highly-relevant use-case of variational quantum approaches, which at the same time has a large potential to reach quantum advantage in the NISQ devices~\cite{Zhang_2022}.

This paper is organized as follows. In Section\,\ref{sec:phenomenology}, we outline the phenomenological theory for the observed algorithmic phase transition and residual infidelity scaling. In Section\,\ref{sec:model_circuits}, we introduce a model for two-dimensional frustrated magnets and symmetry-enhanced circuits. Finally, in Section\,\ref{sec:results} we demonstrate numerical evidence for these claims, and we draw conclusion in Section\,\ref{sec:discussion}.

\section{Phenomenology of state preparation}
\label{sec:phenomenology}

\subsection{An effective stochastic temperature and algorithmic phase transition}
In stochastic optimization methods, such as variational Monte Carlo or gradient-based machine learning, it is known that the power spectrum of statistical noise, under certain assumptions, defines an effective temperature~\cite{Becca_2015,https://doi.org/10.48550/arxiv.1711.04623,NEURIPS2019_dc6a7071}. Concretely, we consider a variational circuit parametrized with a vector of parameters $\boldsymbol{\theta}$ and consider the update law representing the Stochastic Gradient Descent (SGD) approach
$
    \boldsymbol{\theta}^{t + 1} = \boldsymbol{\theta}^t - \eta \nabla \langle \hat H \rangle
$
with $\eta$ being the learning rate. {In the actual multivariate VQE optimization, noise is governed by a $\boldsymbol \theta$--dependent and non-diagonal covariance matrix. Here, to analytically examine the stochastic VQE optimization dynamics,} we assume that $\nabla \langle \hat H \rangle = \mathcal{N}\left(\boldsymbol{f}, \sigma \right),$ i.\,e. that the forces on the parameters, $\boldsymbol{f}_k$, are distributed normally with a diagonal and uniform variance, $\sigma_k^2 \simeq \mbox{Var}\,\boldsymbol{f}_k / N_s$. As mentioned above, in this work, $N_s$ stands for the number of shots used for the estimation of each gradient component.
The effective parameters pseudodynamics is therefore given by a Langevin equation of the form
\begin{gather}
\label{eq:lang}
    \boldsymbol{\theta}^{t + 1}_k = \boldsymbol{\theta}^t_k - \eta \left(\boldsymbol{f}_k + \mathcal{N}(0, \sigma_k) \right),
\end{gather}
where the index $k$ enumerates components of the variational parameters vector of length $N_p$. Besides, we assume that the gradient variances, $\mbox{Var}\,\boldsymbol{f}_k$, are approximately equal for all parameters.\footnote{This assumption is numerically well verified in the case of the Ansatz states considered in the following} Then, stationary solution of Eq.\,\eqref{eq:lang} is the Boltzmann distribution $\Pi(\boldsymbol{\theta}) \propto \exp \left(-E(\boldsymbol{\theta}) / T\right)$, where we defined 
\begin{gather}
    \label{eq:Teff}
    T = \mbox{Var}\,\boldsymbol{f}_k \eta / N_s,
\end{gather}
as an effective temperature of the system. Note that since the learning rate has the dimension of inverse energy, the temperature has the dimension of energy, as expected. This definition agrees with the well-known expression for the effective temperature in variational Monte Carlo optimization~\cite{https://doi.org/10.48550/arxiv.1711.04623,NEURIPS2019_dc6a7071}. 
Without loss of generality, we assume that
the ground state energy value is $E = 0$, and is also reached at $\boldsymbol \theta = \boldsymbol 0$. We thus write to the second order $E(\boldsymbol \theta) = \boldsymbol{\theta}^T \hat{D}\,\boldsymbol{\theta} / 2 = (1/2) \sum_k D_k \tilde{\boldsymbol{\theta}}_k^2$ where we consider a basis $\tilde{\boldsymbol{\theta}}$ delivering a diagonal form to $\hat{D}$~\cite{https://doi.org/10.48550/arxiv.1907.03215,160202666,https://doi.org/10.48550/arxiv.1710.11029}. 
This implies that energy fluctuations are proportional to the number of parameters $N_p$ with $T / 2$ per degree of freedom\footnote{Strictly speaking, the energy fluctuation should be the sum of effective temperatures $\langle E(\boldsymbol \theta) \rangle = (1/2) \sum_k T_k$ associated to the system parameters. In our case, however, for simplicity we assume all temperatures equal.}, namely $\langle E(\boldsymbol \theta) \rangle = (1/2)\,N_p T$.

Below we will provide numerical evidence supporting the existence of an algorithmic phase transition as a function of the energy fluctuations measure $\epsilon = (1/2) N_p T$, separating the $\epsilon > \epsilon^c$ regime where the optimization is impossible from the $\epsilon < \epsilon^c$ regime where the algorithm finds sizable finite overlap with the ground state. Notably, the switch between the two regimes often occurs not through a smooth crossover but is instead characterized by a sharp phase transition with a well-defined $\epsilon^c$, also marked, similar to heat capacity behaviour in second-order phase transitions, by a peak in energy variance $\langle (E - \bar E)^2 \rangle$. We will also show that $\epsilon^c$ does not decrease exponentially with system size, keeping the state preparation feasible while approaching the thermodynamic limit.

We emphasize that our numerical experiment setup is different from the standard thermodynamic case, where both the energy fluctuations and the full problem Hamiltonian energy scale $\Lambda$ are proportional to the number of system degrees of freedom, i.\,e. are both extensive quantities. In our setup, $\Lambda$, being a characteristic of a problem and not of the Ansatz, is independent of $N_p$. In the meantime, energy fluctuations $\epsilon$, depending on the Ansatz, are linear in $N_p$. Intuitively, the ``trainable'' phase should satisfy $\epsilon \ll \Lambda$. Since the latter is independent of $N_p$, the magnitude of energy fluctuations $\epsilon = (1/2) N_p T$, and not temperature $T$, sets a measure detecting the algorithmic phase transition.

Substituting non-diagonal $\boldsymbol \theta$--dependent covariance matrix with diagonal uniform variance $\sigma^2$ is a major approximation. We show, however, that sampling a ``parametric'' partition function
\begin{gather}
\label{eq:Z_thermal}
    \mathcal{Z}_{\boldsymbol \theta} = \int \mbox{d}\boldsymbol{\theta} \exp \left(-E(\boldsymbol \theta) / T \right)
\end{gather}
can reproduce some features of the original stochastic VQE optimization, such as the mentioned parametrical regions with vanishing training fidelity and pronounced separation from the ``trainable phase'' with the transition point marked by a peak of energy variance. Thus, the reported behaviour manifests itself also within sampling the variational circuit parameters from the thermal Boltzmann distribution $\Pi(\boldsymbol \theta) \propto \exp \left(-E(\boldsymbol \theta) / T\right).$

\subsection{A phenomenological scaling law for the residual infidelity}
In the $\epsilon < \epsilon^c$ regime, when the optimization reaches sizable fidelity, we propose the following empirical scaling law
for the residual infidelity $\mathcal{I} = 1 - |\langle \psi(\boldsymbol{\theta})|\psi_{0} \rangle|^2$:
\begin{gather}
    \label{eq:infidelity}
    \mathcal{I} = A \epsilon + \mathcal{I}_0 \, ,
\end{gather}
where $\mathcal{I}_0$ describes the `ideal' representational ineffectiveness of a given variational circuit. It has been demonstrated in an ideal exact gradient setup that $\mathcal{I}_0$ can be made arbitrarily small with a suitable choice of the Ansatz and corresponding circuit depth~\cite{https://doi.org/10.48550/arxiv.2108.08086,Choquette_2021,Mineh_2022}.
However, in addition to the limitations associated with the expressivity of the circuit, the measured state infidelity also stems from the finite sample size of gradient estimates per component, $N_s$.

We argue that this residual infidelity depends on the same effective energy fluctuations measure $\epsilon$ introduced above and on spectral properties of the system of interest, such as the gap to the first excited state $\Delta$. Our numerical experiments suggest a fast scaling with the inverse gap, which to a reasonable degree follows a $A \propto 1 / \Delta^2$ behaviour. 
This proportionality implies an increasing optimization complexity for systems with a closing gap. 
This dependence is similar to the one of the Adiabatic Theorem, which imposes the relation $\tau \propto 1 / \Delta^2$ on the smallest time extent of quantum annealing~\cite{sarandy2005consistency}. In the following, we also provide a recipe allowing one to ameliorate this problem by employing symmetry projections~\cite{Seki_2020,Seki_2022}.

\section{Model and quantum circuits}
\label{sec:model_circuits}
We perform numerical experiments on the $j_1-j_2$ Heisenberg spin--$1/2$ model on a series of two-dimensional lattices.
We place both model and geometries in focus since (i) the ground state of this model under certain conditions realizes a quantum spin liquid, an exotic and long-sought phase of matter~\cite{savary_quantum_2016,zhou_quantum_2017}, (ii) tuning the couplings ratio $j_1/j_2$ allows one, on each of these geometries, to explore different model regimes, and to trigger gap closing, (iii) lattice models allow for a controllable study of the thermodynamic limit.

The spin--$1/2$ Heisenberg model is described by the Hamiltonian
\begin{equation}
    \label{eq:hamiltonian}
    \hat H = j_1 \sum\limits_{\langle i,j \rangle} \hat{\mathbf{S}}_i \cdot \hat{\mathbf{S}}_j + j_2 \sum\limits_{\langle \langle i,j \rangle \rangle} \hat{\mathbf{S}}_i \cdot \hat{\mathbf{S}}_j,
\end{equation}
with $\langle \ldots \rangle$ representing $j_1$--bonds and $\langle \langle \ldots \rangle \rangle$ representing the $j_2$--bonds. The range of ratios $j_2 / j_1$ is chosen, in each particular case, to interpolate between frustrated and magnetically-ordered regimes. See Appendix~\ref{appendix:geometries} for the definition of the lattice geometries and for the description of the parameter regimes for which gap closing is expected.

For numerical experiments with SGD and SR techniques, we employ a symmetry-enhanced Ansatz~\cite{Seki_2020}. The spin-spin interaction in the Hamiltonian (see Eq.\,\eqref{eq:hamiltonian}) can be replaced by the {\it SWAP operator} $\hat P_{ij}~=~\frac{1}{2} \left( \hat{\mathbf{S}}_i \cdot \hat{\mathbf{S}}_j + \hat{\bbone}\right)$, which induces an exchange of spin states between sites $i$ and $j$. If the state is prepared in the total spin-zero $S = 0$ sector, such as the dimer product state
\begin{equation}
\label{eq:dimerized}
|\psi_D\rangle = \bigotimes_{0 \leqslant i < N / 2} \frac{1}{\sqrt{2}} \left(|\uparrow_{2 i} \downarrow_{2i + 1} \rangle - | \downarrow_{2 i} \uparrow_{2i + 1}\rangle \right),
\end{equation}
action of eSWAP operators $\exp \left(i \theta \hat{P}_{ij}\right)$ preserves the total spin and the wave function reads
\begin{equation}
\label{eq:wavefunction}
|\psi(\boldsymbol \theta)\rangle = \left(\prod_{\alpha} e^{i \theta_{\alpha} \hat{P}_{i_{\alpha} j_{\alpha}}} \right) |\psi_D\rangle,
\end{equation}
where to define the $(i_{\alpha},\,j_{\alpha})$ pairs we employ checkerboard decomposition of the Hamiltonian Eq.\,\eqref{eq:hamiltonian}. Specifically, in the case of a square lattice with PBC in both directions, the Hamiltonian can -- up to a constant -- be written as a sum of $L \times 4$ SWAP operators acting between pairs of qubits. We split this set of pairs into $8$ layers of full coverings (when each qubit belongs to exactly one pair), and treat these pairs as $(i_{\alpha},\,j_{\alpha})$. Note that all eSWAP operators within one layer can be applied simultaneously. In addition one can also fix the spatial point group symmetry representation of the wave function (for details of such fixation, optimization protocol and the definition $(i_{\alpha},\,j_{\alpha})$ in the triangular, hexagonal and open boundary condition cases, see Appendix\,\ref{appendix:symmetric_circuit}). 

\begin{figure*}[t!]
    \centering
    \includegraphics[width=\textwidth]{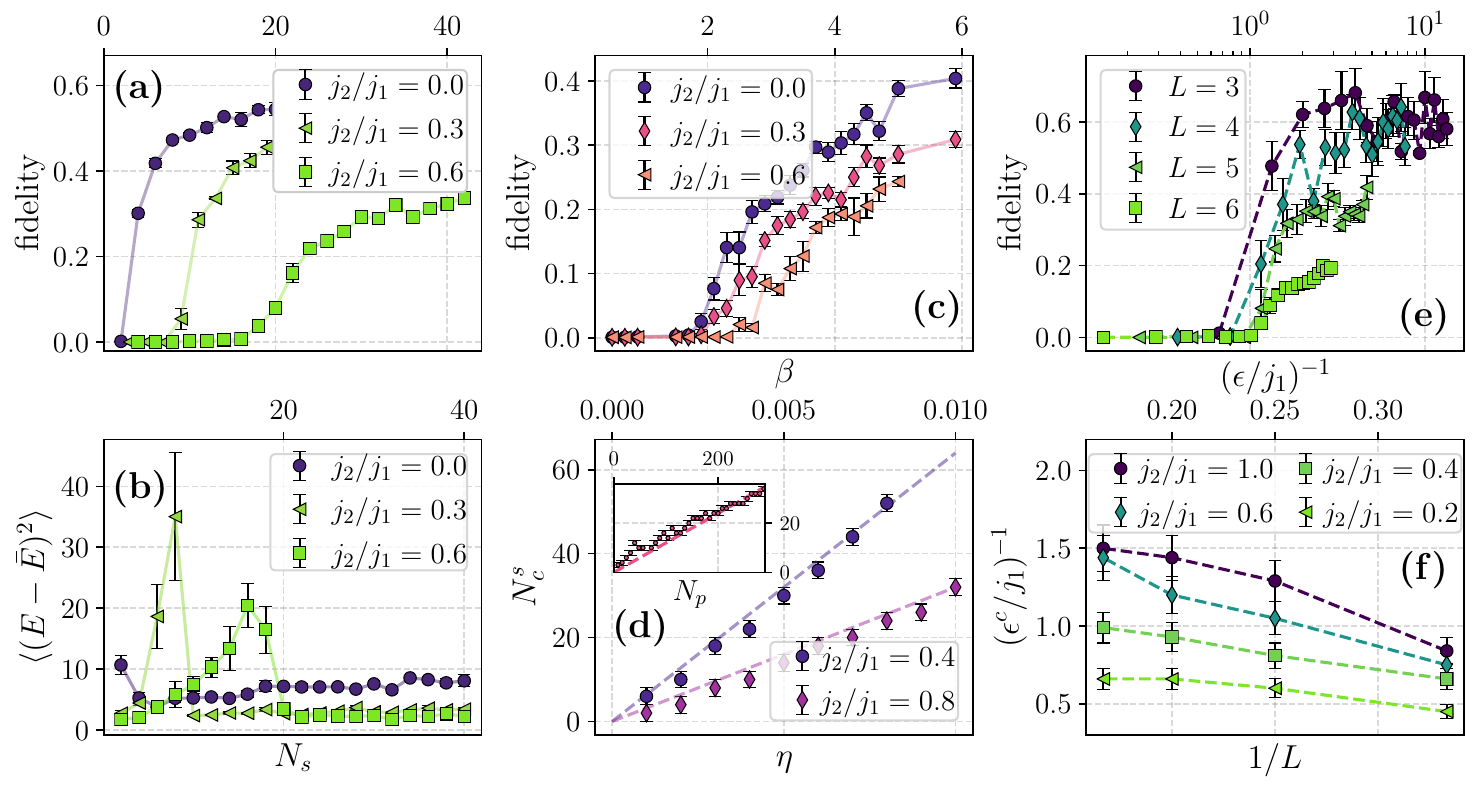}
    \caption{{\bf (a-b)} Fidelity and energy variance as a function of $N_s$ for various $j_2 / j_1$ measured on the $4 \times 4$ square lattice with periodic boundary conditions (PBC).\; {\bf (c)} Fidelity as a function of inverse temperature $\beta$ within simulation of the $4 \times 4$ square lattice at various $j_2 / j_1$ with direct sampling from thermal partition function Eq.\,\eqref{eq:Z_thermal}. {\bf (d)} Critical number of samples $N_s^c$ as a function of learning rate $\eta$ on the $4 \times 4$ triangular lattice with PBC. (d, inset): Critical number of samples $N_s^c$ as a function of number of parameters $N_p$ on the $4 \times 4$ square lattice with PBC at $j_2 / j_1 = 0.4$. {\bf (e-f)} Circuit fidelity as a function of $\epsilon / j_1$ within the $L \times 4$ setup at $j_2 / j_1 = 0.4$ and inverse critical fluctuation measure $(\epsilon^c / j_1)^{-1}$ as a function of $1 / L$.}
    \label{fig:panel1}
\end{figure*}

This Ansatz allows one to effectively change $\Delta$ (if symmetry is imposed, effective $\Delta$ is the energy of the first excited state {\it in the respective irreducible representation}) and improves the ability to express the ground state (decreases $\mathcal{I}_0$), keeping $N_p$ relatively low.
Importantly, the introduced Ansatz only shows a mild decay of the gradient norms with system size, signaling the absence of the so-called barren plateau issue, proven to emerge in a generic quantum circuit setup~\cite{McClean_2018}. 
The proposed circuit has a sub-exponential decay of gradients, similarly to the case of the transverse field Ising model reported in~\cite{PRXQuantum.1.020319}, as we demonstrate in Appendix~\ref{appendix:absence_of_barren_plateaus}. Other barren plateaus-free circuits aimed at the study of frustrated two-dimensional magnets were also previously reported~\cite{https://doi.org/10.48550/arxiv.2108.08086,Liu_2019}. Finally, we note that the eSWAP gates are simply the restriction of the two-qubit rotation gates $\hat{U}_{ij}(\theta, \varphi) = \exp \left(i \theta (\hat{X}_i\hat{X}_j + \hat{Y}_i\hat{Y}_j) + i\varphi \hat{Z}_i \hat{Z}_j\right)$ to the case $SU(2)$--symmetric case $\varphi = \theta$. These two-qubit gates are the standard choice for the VQE studies of spin--$1/2$ Heisenberg magnets~\cite{Seki_2020,Jattana_2022,https://doi.org/10.48550/arxiv.2205.11198,https://doi.org/10.48550/arxiv.2108.08086,https://doi.org/10.48550/arxiv.2108.02175}. Therefore, the proposed ansatz Eq.\,\eqref{eq:wavefunction} represents a rather general family of circuits with the additional advantage of imposed spin quantum number conservation.

\section{Results}
\label{sec:results}
\subsection{Small sample size regime: critical behaviour}
The optimization of the circuit parameters requires the estimation of the energy gradient with respect to gate parameters $\boldsymbol \theta$, for which a circuit is being executed $N_s$ times per gradient component. Therefore, the total number of circuit shots is $N_s N_{\text{par}}.$ Larger $N_s$ leads to better estimates for the wave function projections such as the fidelity $\mathcal{F} = |\langle \psi(\boldsymbol \theta) | \psi_0\rangle|^2$ (i.\,e., the squared overlap with the ground state), as well as for expectation values of the form $\langle \psi(\boldsymbol \theta) |\hat O| \psi(\boldsymbol \theta) \rangle$ for any given observable $\hat O$. A naively expected behavior in such a case would be a gradual growth of fidelity with $N_s$. However, Ref.\,\cite{2020_twesterhout}, which studies Neural Quantum States (NQS)\,\cite{2017_gcarleo} (an ansatz class used in \textit{classical} computing) reported a sharply different behavior with nearly zero fidelity for $N_s < N_s^c$ smaller than some critical $N_s^c$ number of samples\footnote{Another prominent  example of an algorithmic phase transition occurring in an optimization problem is the so-called {\it jamming transition} reported for an artificial Neural Network training~\cite{Spigler_2019}.}.

\subsubsection{Manifestation of the critical behavior}
To test which scenario (a smooth transition or a critical behavior) is realized with the quantum state preparation, we perform optimization with stochastic gradient descent (SGD) in the small--$N_s$ regime\footnote{In this regime of small $N_s$, the metric tensor inversion requires a very large regularization (see Appendix\,\ref{appendix:symmetric_circuit}), effectively turning SR into SGD.}. We plot the circuit fidelity (or, similarly, infidelity) throughout the Results section on the vertical axis. To obtain the circuit fidelity, we employ the following protocol. Starting from a random set of parameters, we optimize a circuit using the SGD or SR approach until the sliding mean of fidelity with a window of several hundred steps stabilizes. 

After such convergence, we use the mean over the next thousand iteration steps as the circuit fidelity. The procedure is repeated ten times to estimate the error bars. In Fig.\,\ref{fig:panel1}\,(a) we show the state fidelity as a function of $N_s$ for various $j_2 / j_1$ on the $4 \times 4$ square lattice with open boundary conditions (OBC) in the first and periodic (PBC) in the second dimension to avoid geometrical frustrations in the $j_2 / j_1 = 0$ pure nearest-neighbor case. One can see that the fidelity remains vanishingly small for $N_s < N_s^c$, followed by a rapid growth afterwards. Such drastic change of pattern is suggestive of a {\it critical behaviour} (see Appendix\,\ref{appendix:distribution_overlaps} for similar behavior in other frustrated lattices). 

To explore this transition in $N_s$, in Fig.\,\ref{fig:panel1}\,(b) we show $c_V / N_s^2 = \mbox{Var}\,E$ as a function of $N_s$, where $c_V$ is the effective specific heat. We observe that the peak of $\mbox{Var}\,E$ coincides precisely with $N_s^c$, defined as the departure point from the zero-fidelity regime. 
The effective specific heat $c_V$ also shows a peak in this region. This quantity generalizes the thermal specific heat $c^{\beta}_V = \beta^2\, \mbox{Var}\, E$, where we assume that the inverse number of samples, $1 / N_s$, plays the role of an effective temperature, and $\beta$ is the inverse temperature. We observe a similar critical behaviour across other cluster dimensions and geometries under consideration. Notably, the reported algorithmic phase transition is also accompanied by a qualitative change in the distribution of the overlaps with higher excited states $\mathcal{O}_k = |\langle \psi(\boldsymbol{\theta}) | \psi_k \rangle|^2$. Namely, it changes from being peaked at $1 / |\mathcal{H}|$ (with $|\mathcal{H}|$ being the size of the Hilbert space) to be much wider (see Appendix.\,\ref{appendix:distribution_overlaps}). Thus, at small $N_s$, the circuit learns {\it uniform} overlap with all eigenstates, while at $N_s > N_s^c$ the circuit favors only several low-lying excitations. This is consistent with the argument outlined in Section.\,\ref{sec:phenomenology}, based on the thermal partition function.

\begin{figure*}[t!]
    \centering
    \includegraphics[width=\textwidth]{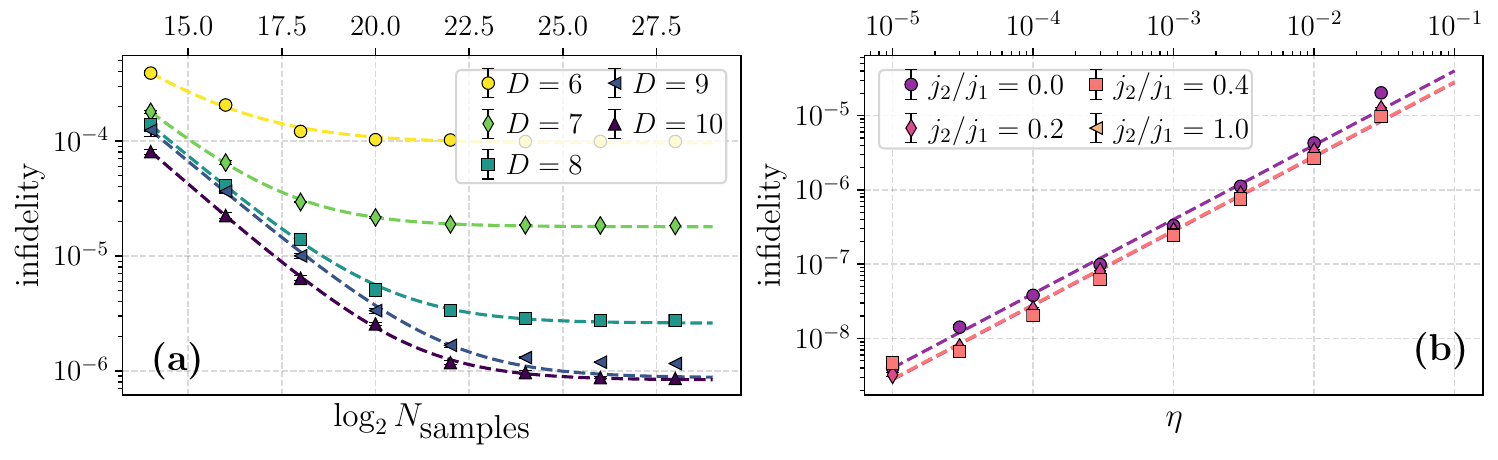}
    \caption{Large-$N_s$ regime study of the $4 \times 4$ square lattice. {\bf (a)} Infidelity as a function of $N_s$ for a set of depths $D = N_p / 8$ at $j_2 / j_1 = 0.4$ computed with $\eta = 10^{-2}.$ The dashed line show fit with the Ansatz Eq.\,\eqref{eq:infidelity}. {\bf (b)} Infidelity as a function of $\eta$ for a set of $j_2 / j_1$ at $D = 8$. The data was generated at $N_s = 2^{20}$ to ensure convergence and well-conditioning of the metric tensor at any value of $\eta$ under consideration.}
    \label{fig:panel2}
\end{figure*}

To substantiate this, we simulate the ``parametric'' partition function $\mathcal{Z}_{\boldsymbol \theta}$ defined in Eq.\,\eqref{eq:Z_thermal} using a Metropolis algorithm that performs random walks in the parameter space.
In Fig.\,\ref{fig:panel1}\,(c) we show the average fidelity computed along the generated Markov chain at different temperatures for the $4 \times 4$ square lattice. We keep the number of parameters fixed and thus present the data as a function of $\beta = 1/T$. Even though the growth after the critical value $\beta_c$ is not as sharp as observed in Fig.\,\ref{fig:panel1}\,(a), we notice also in this case a region of nearly zero fidelity for $\beta < \beta_c$. Similarly to the case in Fig.\,\ref{fig:panel1}\,(b), the specific heat $c^{\beta}_V$ has a pronounced peak at $\beta_c$. The histogram of overlaps $\mathcal{O}_k$ shows a similar qualitative change between the two regimes (see Appendix\,\ref{appendix:distribution_overlaps} for details).

We note that the abruptness of growth at $N_s > N_s^c$ ($\beta > \beta_c$) is clearly dependent on $j_2 / j_1$, as seen from contrasting the $j_2 / j_1 = 0.6$ and $j_2 / j_1 = 0.0$ curves demonstrated in Fig.\,\ref{fig:panel1}\,(a, c). However, the transition at $j_2 / j_1 = 0$ appears at $N_s \to 0$, but at finite $\beta$. We believe, this is due to the sampling from the thermal partition function and neglecting correlation between the noise components. Nevertheless, clear separation between the two algorithmic regimes marked by a peak of energy variance, remains present.

To verify the effective temperature expression Eq.\,\eqref{eq:Teff}, in Fig\,\ref{fig:panel1}\,(d), we plot $N_s^c$ as a function of $\eta$ for the $4 \times 4$ triangular lattice. To estimate the transition position $N_s^c$, we take the smallest $N_s$ such that the averaged fidelity exceeds double inverse Hilbert space size. From Fig\,\ref{fig:panel1}\,(d), we observe a linear dependence on learning rate $\eta$. This supports the linear $\eta$ proportionality in Eq.\,\eqref{eq:Teff}. The inset shows $N_s^c$ as a function of $N_p$. As expected, we observe a linear dependence, while the variance per parameter, $\frac{1}{N_p} \sum_k \mbox{Var}\,\boldsymbol{f}_k$, remains mostly unchanged. This behavior is also observed in other lattice geometries. This shows a pronounced criticality in the energy fluctuations measure $\epsilon = (1/2) T N_p$, rather than in the mere effective temperature $T$. 

\subsubsection{Critical behavior in the thermodynamic limit}

We aim to extrapolate the critical energy fluctuations measure $\epsilon^c$ to the thermodynamic limit. To this end, we consider $L \times 4$ square lattices with $L = 3,\,4,\,5$ and $6$. As circuit architecture, we select the 8 checkerboard decomposition layers of the Hamiltonian in  Eq.\,\eqref{eq:hamiltonian}. Thus, due to the high computational cost otherwise, the depth of the circuit does not scale with $L$.
Additionally, OBC in the $L$--direction and PBC in the second direction are imposed, ensuring the absence of geometric frustrations. In Fig.\,\ref{fig:panel1}\,(e) we report the state fidelity as a function of $\epsilon$, while in Fig.\,\ref{fig:panel1}\,(f), we show the dependence of $\epsilon^c$ on the lattice dimension $L^{-1}$. We observe a saturation with the increase of $L$, and note that the critical energy fluctuations do now grow exponentially with the system size\footnote{We note that $\epsilon_c / j_1$ varies only mildly with $j_2 / j_1$, while the system gap has a two orders of magnitude difference between $j_2 / j_1 = 0.0$ and $0.6$. This suggests that in the trainable phase definition $\epsilon \ll \Lambda$ given above, $\Lambda$ is not an energy gap of $\hat H$, but rather some other energy scale of the system.}. Note, that the curves in Fig.\,\ref{fig:panel1}\,(e) approach different final asymptotic values $\mathcal{F}_{\infty}$ for $\epsilon < \epsilon^c$, while ``fair'' extrapolation to thermodynamic limit should ensure non-vanishing (preferably same) saturation. {Thus, as $L$ increases, one needs to employ deeper circuits, which would correspondingly increase $\epsilon^c$ at a given $L$. However, due to high computational cost, we refrain from scaling up the circuit depth at large cluster volumes $L = 5,\,6$. Nevertheless, we believe that such an increase in the circuit depth would not change the overall sub-exponential scaling character of $\epsilon^c(L)$. For instance, Ref.\,\cite{bosse2021probing} reports that, in order to reach a certain fidelity $\mathcal{F}$ with the ground state on the frustrated $N$--site kagome lattice, only $\propto \sqrt{N}$ gates are necessary. Thus, with such depth adjustment, saturating behavior of $\epsilon^c(L)$ shown in Fig.\,\ref{fig:panel1}\,(f) can only change to polynomial dependence on $L$, and not to exponential growth.}

\begin{figure*}[t!]
    \centering
    \includegraphics[width=\textwidth]{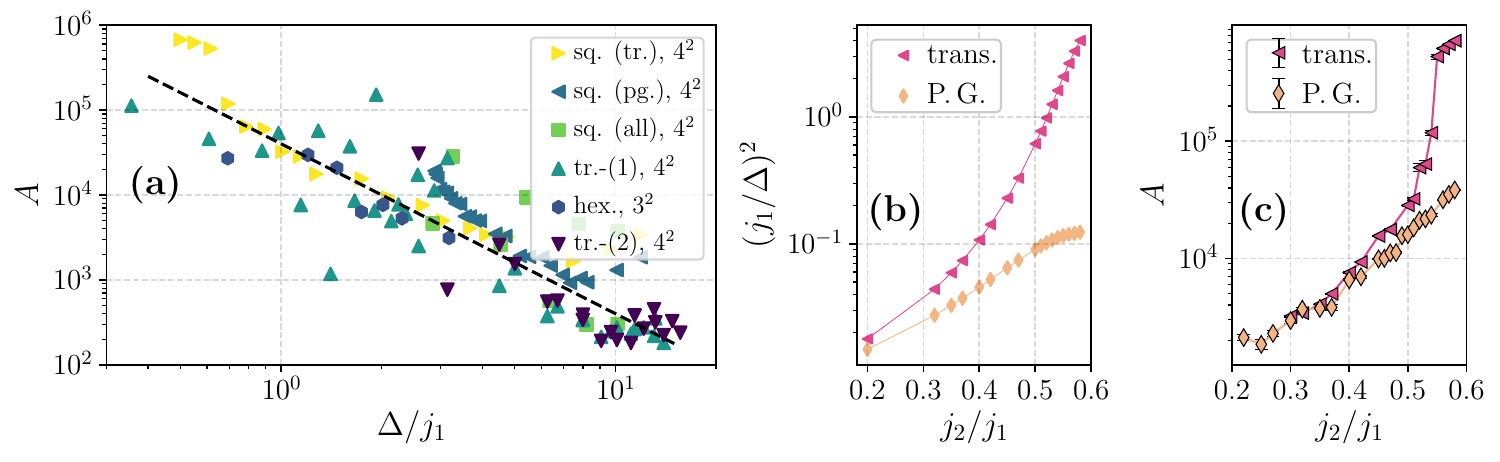}
    \caption{{\bf (a)} Dependence of the prefactor $A$ introduced in Eq.\,\eqref{eq:infidelity} on the system spectral gap. The symbols denote: {\it sq. (tr.), $4^2$} --- square $4^2$ lattice case with only translational symmetry imposed, {\it sq. (pg.), $4^2$} --- same with only point symmetry group imposed, {\it sq. (all), $4^2$} --- same with both translation and point group symmetries imposed, {\it tr.-(1), $4^2$} --- triangular $4^2$ lattice with full symmetry imposed with starting dimerization along the $j_1$--bonds, {\it tr.-(2)} --- same with dimerization along the $j_2$--bonds, and {\it hex., $3^2$} --- $3^2$ hexagonal lattice with full symmetry group imposed. {\bf (b)} On the $4 \times 4$ square lattice, dependence the system gap $(j_1 / \Delta)^2$ on $j_2 / j_1$ in two cases: (i) with only point group symmetry imposed, (ii) with only translational symmetry imposed. {\bf (c)}: same for the resulting prefactor $A$.}
    \label{fig:panel3}
\end{figure*}

The main practical quantity of interest is $N_{\text{tot}}(L) = N_{\text{SGD}}(N_s^c) \times N_s^c$, the total number of samples required to train a circuit in the course of $N_{\text{SGD}}(N_s^c)$ SGD iterations, which is proportional to the required hardware resources. We define $N_{\text{SGD}}(N_s)$ as the number of SGD steps needed to reach 90\,\% of the saturated fidelity. 
We observe that $N_{\text{SGD}}(N_s)$ required to saturate $\mathcal{F}_{\infty}$ (as defined in Fig.\,\ref{fig:panel1}\,(e)) shows no growth with $L$ (see Appendix.\,\ref{appendix:N_SGD} for details).
Thus, the obtained scaling for $\epsilon^c(L)$ results into polynomial scaling of $N_{\text{tot}}(L)$. Note that reducing $\eta$ will allow reaching $\epsilon^c$ with a smaller number of samples per SGD iteration. However, such convergence would require proportionally more SGD steps $N_{\text{SGD}}$, making $N_{\text{tot}}(L)$ a true lower bound on the required total number of samples.

\subsection{Large sample size regime}
In this section, we provide numerical evidence to validate Eq.\,\eqref{eq:infidelity} in the regime of accurate gradients. We employ the symmetrized wave function, and the Stochastic Reconfiguration (SR) approach proposed in~\cite{Sorella_1998}. In this section, $N_s$ has the meaning of the number of shots per gradient component and per symmetry projection (see Appendix\,\ref{appendix:symmetric_circuit} for projections definition).

We consider the $4 \times 4$ square lattice at $j_2 / j_1 = 0.4$ with a variable circuit depth. The circuit is built of repetitive applications of the checkerboard decomposition of the Hamiltonian in Eq.\,\eqref{eq:hamiltonian} (a single decomposition is described in Section\,\ref{sec:model_circuits}), restricting ourselves to a maximum of $N_p$ optimization parameters. The circuit depth is then $D = N_p / 8$. In Fig.\,\ref{fig:panel2}\,(a) we present the infidelity as a function of $N_s$. To verify the proposed functional form for $\mathcal{I}(N_s)$, the data are fitted with the expression in Eq.\,\eqref{eq:infidelity}. To verify the power law, we replace $N_s \to N_s^{\alpha}$ and observe $\alpha \sim 1$ within error bars consistently within the fitting procedure (for details, see Appendix\,\ref{appendix:fit_parameters}). Finally, in Fig.\,\ref{fig:panel2}\,(b) we fit infidelity with a linear function $C \eta$, where $C$ is a fit constant. These fits support the functional form Eq.\,\eqref{eq:infidelity} and provide an empirical way to verify the $1 / N_s$ power law dependence of the residual infidelity.

In the following, we investigate the dependence of the prefactor $A$ on the spectral gap $\Delta$. To this end, in Fig.\,\ref{fig:panel3}\,(a) we present the offset $A$ against $(\Delta / j_1)^{-1}$ for the different two-dimensional lattices considered in this study. We consider different lattice geometries, different starting dimerization patterns, and symmetry projectors. The fit with $A_0 \Delta^{\alpha}$ yields $\alpha \sim -2$. The black dashed line shows the $A_0 / \Delta^2$ fit of the data. {We note that in the cases of other spin systems that we considered, such as the $j_1$--$j_2$ Heisenberg chain, we observed a scaling law  significantly different from $1 / \Delta^2$ and being highly depth-dependent. We thus emphasize that our claim $\alpha \sim -2$ only concerns two-dimensional frustrated magnets.}

From this observation, it follows that for increasingly smaller gaps, the systems become intractable, in the sense that it becomes harder to converge to the true ground state. However, imposing symmetrization of the Ansatz (see Appendix.\,\ref{appendix:symmetric_circuit}) can alleviate this problem. We illustrate this approach in the case of the $4 \times 4$ square lattice, where the parameter regime $j_2 / j_1 \to 0.6$ leads to a nearly closing gap, accompanied by the emergence of a quantum spin liquid~\cite{Nomura_2021}. Namely, we contrast the cases of (i) only point group symmetry imposed with $8$ terms in the projector against (ii) only translations imposed with $16$ terms in the projector. As seen in Fig.\,\ref{fig:panel3}\,(b) at $j_2 / j_1 \to 0.6$, $(j_1 / \Delta)^2$ diverges for case (ii), while it remains well-behaving for case (i). This results in dramatically different behaviour for $A(j_2 / j_1)$, with the latter being moderate for (i) and exploding for (ii), as seen in Fig.\,\ref{fig:panel3}\,(c). This improvement in the prefactor $A$ magnitude emphasizes the importance of symmetries in the state preparation process of systems with a vanishing gap. In summarizing, the residual state infidelity, which is not due to the lack of representability of the chosen Ansatz, but rather to the optimization process, scales as
\begin{gather}
    \label{eq:infidelity_gap}
    \mathcal{I}-\mathcal{I}_0 \propto  {\epsilon \over \Delta^2}.
\end{gather}
where $\epsilon = (1/2) T N_p$ with $T$ defined in terms of sampling shots $N_s$, and learning rate $\eta$, as in Eq.\,\eqref{eq:Teff}.

\section{Discussion}
\label{sec:discussion}
Variational state preparation is essential for various quantum computing algorithms in near-term and fault-tolerant regimes.
Here we show that, besides the most apparent dependency on the circuit architecture, the fidelity of the prepared state strongly depends on learning hyperparameters and the system-dependent properties. 
We layout a phenomenology of state preparation as a function of two significant factors, (i)
the number of samples $N_s$ used to estimate the gradient of the variational parameters in an SGD step, and (ii) the fundamental gap $\Delta$ of the model Hamiltonian.
In particular, we explored the interplay of these two parameters, focusing on challenging two-dimensional spin-$1/2$ frustrated quantum magnets. These problems represent the most challenging cases for the classical approaches, while retain locality in the qubit form, unlike quantum chemistry formulations. These problems are therefore marked as the most promising candidates to reach quantum advantage in the variational quantum studies. To express the ground state of such model, we employ the $SU(2)$--protecting circuit consisting of eSWAP gates, which is only a mild restriction of the XXZ two-qubit rotations, standard in the variational studies of the spin-$1/2$ Heisenberg models. This circuit is optimized using the SGD approach, a scalable optimization method in the limits of large parameters number. In this rather general setup, we report novel challenges for the VQE optimization.

Namely, we observe that, in the regime of small $N_s$ (noisy gradient) the stochastic optimization shows critical behaviour with near-zero state fidelity for $N_s < N_s^c$ and rapid growth of fidelity at $N_s > N_s^c$. The pace of fidelity growth is highly geometry- and parameter regime-dependent. However, the zero-fidelity region is always present and is pronouncedly separate from the ``trainable phase''. Besides, the point of transition, $N_s^c$, always features a peak of energy variance $\langle (E - \bar E)^2 \rangle$, resembling heat capacity behaviour in second-order phase transitions. Together with the notion of effective temperature, this separation allows us to discuss an effective algorithmic phase transition on the energy fluctuations measure $\epsilon = (1/2) N_p T$ axis. In the case of a two-dimensional square lattice, we found evidence that the critical energy fluctuation $\epsilon_c$ scales only polynomially with the system size (see Fig.\,\ref{fig:panel1}\,(f)), providing the basis for possible applications of VQE in the study of larger-size frustrated magnets, inaccessible to classical algorithms.

To support the notion of effective temperature, we show that the observed criticality and energy variance peak can be reproduced within sampling from a simpler ``parametric'' partition function Eq.\,\eqref{eq:Z_thermal} with the classical circuit parameters $\boldsymbol \theta$ distributed according to the Boltzmann weight.
We thus provide a simplified picture explaining the reported algorithmic phase transition in $\epsilon$ and justifying the notion of an effective temperature of VQE optimization.

The observed threshold $N_s^c$ sets a minimum non-negligible
runtime of a successful VQE algorithm on realistic future quantum hardware. To see this, we consider a small yet interesting case of a $N_q = 10 \times 10$ square lattice. The current reasonable estimation for a single eSWAP gate runtime is in the order of $10^{-4}\,\text{s}$ in the fault-tolerant quantum regime~\cite{https://doi.org/10.48550/arxiv.1808.06709}. With the depth of the circuit of order $N_q$, a single shot implies a walltime  $\tau = 0.01\,\text{s}.$ Therefore, the total run time of each SGD iteration at $N_s = N_s^c$, is of order $\tau \times N_s^c \times N_{\text{par}} = 0.01\,\text{s} \times 20 \times 5000 \sim 10\,\text{min}$. Here, we estimated the number of parameters as $5000 = \text{depth} \times (50 \text{ eSWAP dimer pairs on each layer})$, and $N_s = 20$ per gradient component. Of course, translationally-invariant architectures can be employed, to reduce $N_{\text{par}}$ or circuit depth~\cite{https://doi.org/10.48550/arxiv.2009.09423}. 
Nevertheless, our finding suggests the existence of a non-vanishing minutes-order lower bound of quantum processing time (QPU) per optimization step in a potential case of applying VQE to a classically-hard system.

We have also observed that the approach to the exact ground state in the large--$N_s$ limit depends heavily on the system spectral properties. Namely, in the studies case of two-dimensional frustrated magnets, the $N_s$--dependent contribution to the residual infidelity Eq.\,\eqref{eq:infidelity} scales as $1 / \Delta^2$ with $\Delta$ being the energy gap of the studied system. We emphasize that this empirical dependence is only applicable to two-dimensional magnets that were explored in this work. Moreover, even though we express $A$ as the function of the gap to the first excited state, $\Delta$, the $1/\Delta^2$ scaling is also related to the growing contributions to the infidelity coming from the higher-excited states. Namely, as we show in Appendix\,\ref{appendix:overlaps_distribution} that approaching the gap-closing point on the $j_2 / j_1$ axis is accompanied by the growing low-energy excited states density. The interplay between these two factors might lead to the observed $A$ scaling.

We note that the observed $1/\Delta^2$ infidelity scaling is fundamentally different from the well-known runtime scaling coincidentally appearing in adiabatic quantum computing~\cite{PhysRevE.58.5355} and incoherent quantum tunneling~\cite{PhysRevB.35.9535,PhysRevLett.117.180402,PhysRevB.96.134305}. Moreover, a simple two-level model would yield a $1 / \Delta$ scaling, also observed in Ref.\,\cite{McClean_2016} (Eq.\,(25)). This is because, if only the first excited state plays a role, then $\langle E \rangle = \Delta \langle \mathcal{I}\rangle$. Therefore, since we have shown that $\langle E \rangle \sim 1/T \sim 1/N_s$, we arrive at $\langle \mathcal{I}\rangle \sim 1/(\Delta N_s),$ which yields the $1/\Delta$ scaling of $N_s$ with the gap. In contrast, in the case of the systems considered in this work, higher excited states also contribute, which leads to a ``worse'', $1 / \Delta^2$, scaling.

This $1 / \Delta^2$ scaling poses a significant obstacle for the study of systems with a closing gap. To address this problem, we showed how a symmetry-enhanced wave function -- in addition to being not susceptible to the barren plateaus issue -- can, in some cases, mitigate the effects of a closing gap. Namely, imposing symmetry projection can increase the effective system gap $\Delta$ in this symmetry sector, relieving the exploding $A = 1/\Delta^2$ prefactor and providing a significant improvement in the algorithm efficiency, which can be quantified in several orders of magnitude. This development will open up new possibilities for using VQE to simulate complex many-body systems with near-term quantum computers, which could otherwise be intractable due to a closing gap.

\section{Acknowledgements}
We are sincerely grateful to T.\,Westerhout for useful discussions and help with the technical simulation setup. We thank Titus Neupert for helpful comments on our manuscript. Numerical simulations used the high-performance package \texttt{lattice{\_}symmetries}~\cite{westerhout2021latticesymmetries} for quantum state vectors manipulation. N.\,A is funded by the Swiss National Science Foundation, grant number: PP00P2{\_}176877. N.\,A. acknowledges the usage of computing resources of the federal collective usage center ``Complex for simulation and data processing for mega-science facilities'' at NRC ``Kurchatov Institute''.

\section{Appendix}
\subsection{Two-dimensional geometries}
\label{appendix:geometries}
\begin{figure}[h!]
    \centering
    \includegraphics[width=0.49\textwidth]{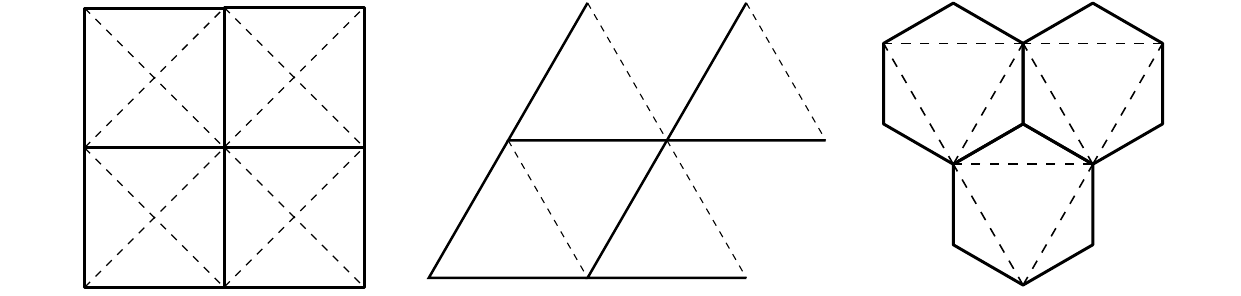}
    \caption{Geometries of lattices considered in this work. Solid lines represent $j_1$--bonds, while $j_2$ bonds are represented by dashed lines: {\bf (a)} square lattice, {\bf (b)} triangular lattice, {\bf (c)} hexagonal lattice. In the cases of $L \times L$ equilateral $4 \times 4$ triangle and $3 \times 3$ kagome lattices considered in this work, $L$ is the number of 1-site (triangle) and 3-site (kagome) unit cells in each lattice dimension. In the case of the $L \times 4$ square lattice, $L$ is the number of unit cells in the $x$-direction.}
    \label{fig:geometries}
\end{figure}

In Fig.\,\ref{fig:geometries} we show two-dimensional lattice geometries considered in this work: square, triangle, and hexagonal. The $j_1$ bonds are marked as solid lines, while the $j_2$ bonds are shown as dashed lines. Each lattice shows a near-vanishing gap to the first excited state in the $S = 0$ sector in some region of $j_2 / j_1$. For instance, in the case of a square lattice, the vicinity of the $j_2 / j_1 \approx 0.55$ fluctuations melt magnetic orders and result in a frustrated phase, possibly gapless QSL~\cite{Nomura_2021}. On the triangular lattice, gap shrinks in the vicinity of $j_2 / j_1 = 1.0$, where the model is most frustrated~\cite{Iqbal_2016} and in the vicinity of $j_2 / j_1 = 0.2$ on the hexagonal lattice~\cite{PhysRevB.88.165138}.

\subsection{Symmetric circuit routines}
\label{appendix:symmetric_circuit}
\begin{figure}[t!]
\centering
    \includegraphics[width=0.49\textwidth]{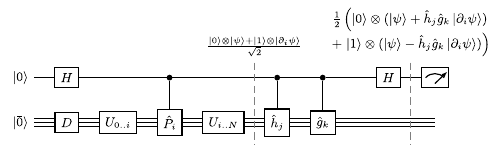}

\caption{Circuit for measurement of $\langle \psi | \hat h_j \hat g_k | \partial_i \psi \rangle$. The Hadamard scheme applied to the ancilla qubit allows to obtain real and (if necessary for connection and metric tensor) imaginary part of $\langle \psi | \hat h_j \hat g_k | \partial_i \psi \rangle$}
\label{fig:circuit}
\end{figure}

\begin{figure}[b!]
\centering
    \includegraphics[width=0.49\textwidth]{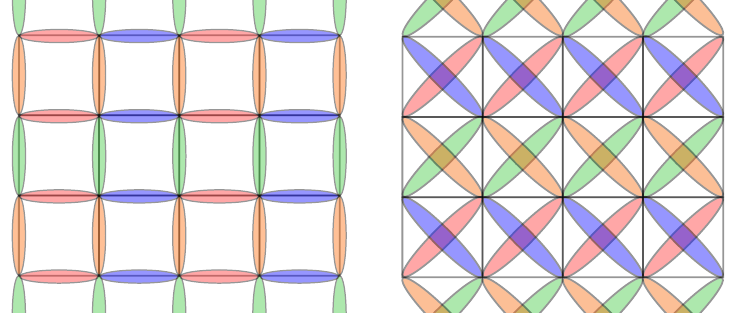}

\caption{Decomposition of the Hamiltonian terms into 8 layers of full dimerization in the case of the $5 \times 4$ square lattice with PBC (vertical) and OBC (horizontal) boundary conditions.}
\label{fig:appendix_ham_split}
\end{figure}

{\it Symmetrized wave function:} In the Heisenberg Hamiltonian, spin-spin interaction can be replaced with the {\it SWAP operator} $\hat P_{ij}~=~\frac{1}{2} \left( \hat{\mathbf{S}}_i \cdot \hat{\mathbf{S}}_j + \hat{\bbone}\right)$, exchanging spin states on the site $i$ and $j$. Notably, the SWAP operator commutes with the total spin operator $[\hat P_{ij}, \hat{S}^2] = 0$, allowing for working in the wave function sector with fixed total spin~\cite{Seki_2020}. As the result, the symmetry-enhanced Ansatz for the system of $N$ spins is constructed by (1) first preparing the system in the simple {\it fully-dimerized} state
\begin{equation}
\label{eq:dimerized}
|\psi_D\rangle = \bigotimes_{0 \leqslant i < N / 2} \frac{1}{\sqrt{2}} \left(|\uparrow_{2 i} \downarrow_{2i + 1} \rangle - | \downarrow_{2 i} \uparrow_{2i + 1}\rangle \right),
\end{equation}
being the direct product of $N / 2$ dimer pairs\footnote{The exact dimerization pattern is chosen to maximize overlap with the ground state} and (2) action of the string of parametrized eSWAP operators preserving the total spin $S = 0$\footnote{To construct wave function in another spin sector, one needs to prepare one (or more) electron pairs in spin-triplet state $|\uparrow \uparrow\rangle$.}:
\begin{equation}
\label{eq:wavefunction}
|\psi\rangle(\boldsymbol \theta) = \left(\prod_{\alpha} e^{i \theta_{\alpha} \hat{P}_{i_{\alpha} j_{\alpha}}} \right) |\psi_D\rangle.
\end{equation}

To define $i_{\alpha}$ and $j_{\alpha}$, we employ checkerboard decomposition of the Hamiltonian Eq.\,\eqref{eq:hamiltonian}. Namely, in case of a square lattice, we split the Hamiltonian terms into $8$ layers of full dimerizations, when all eSWAP operators within one layer can be applied simultaneously\footnote{In the triangle lattice case, to construct the ($i_{\alpha}$, $j_{\alpha}$) pairs, we effectively add the ``missing'' $j_2$ bonds (with zero exchange interaction) to recover the square lattice setup and employ the same decomposition. In the case of the $3 \times 3$ hexagonal lattice, the $j_1$ and $j_2$ bonds groups are split into incomplete dimerization coverings (including no all spins at once) with parallel dimers.}. In this work, we consider OBC-PBC and PBC-PBC boundary conditions. For instance, in the case of the $4 \times 5$ square lattice, the partition of the Hamiltonian terms into 8 layers is shown in Fig.\,\ref{fig:appendix_ham_split}. The spatial symmetry projector operator is defined as 
\begin{gather}
    \label{eq:symmetrization}
    \hat P = \frac{1}{|G|} \sum\limits_{g \in G} \chi_g \hat g,
\end{gather}
where $G$ is the spatial symmetry group, consisting of the elementary unitary permutations $\hat g$ and $\chi_g$ are the characters, depending on the desired projection quantum number. The projected wave function  $|\psi_P(\boldsymbol \theta)\rangle = \frac{\hat P}{\sqrt{\mathcal{N}(\boldsymbol \theta)}} |\psi(\boldsymbol \theta)\rangle$ is normalized with $\mathcal{N}(\boldsymbol \theta) = \langle \psi(\boldsymbol\theta) | \hat P | \psi(\boldsymbol \theta) \rangle$. The energy gradient reads
\begin{gather}
    \label{eq:gradient}
    \partial_i \langle E(\boldsymbol \theta) \rangle = 2 \mbox{ Re} \left[ \frac{\langle \psi(\boldsymbol \theta) | \hat H \hat P | \partial_i \psi(\boldsymbol \theta)\rangle}{\mathcal{N}(\boldsymbol\theta)} - \mathcal{A}_i(\boldsymbol \theta) \langle E(\boldsymbol \theta) \rangle \right],
\end{gather}
where $\mathcal{A}_i(\boldsymbol \theta) = \frac{1}{\mathcal{N}(\boldsymbol \theta)} \langle \psi(\boldsymbol \theta) | \hat P | \partial_i \psi(\boldsymbol \theta) \rangle$ is the {\it connection}.

Finally, the metric tensor for natural gradient descent is defined as 
\begin{equation}
\label{eq:metric_tensor}
    G(\boldsymbol \theta)_{ij} = \frac{\langle \partial_i\psi(\boldsymbol \theta) | \hat P | \partial_j \psi(\boldsymbol \theta)\rangle}{\mathcal{N}(\boldsymbol\theta)} - \mathcal{A}^*_i(\boldsymbol \theta) \mathcal{A}_j(\boldsymbol \theta),
\end{equation}
and used to improve energy gradient $\boldsymbol \theta_{k + 1}=\boldsymbol \theta_k - \eta \sum\limits_j \left(\mbox{Re}\, G(\boldsymbol \theta)  \right)^{-1}_{ij} \partial_j \langle E(\boldsymbol \theta) \rangle$ in the spirit of imaginary time evolution within stochastic reconfiguration (SR)~\cite{Sorella_1998}. The metric tensor obtained within sampling is regularized $G_{\text{reg}} = \sqrt{G G} + \beta \hat{\bbone}$ as suggested in Ref.\,\cite{gacon2021simultaneous}.

{\it Sample quantum circuit:} In the course of optimization, it is required to measure expectation values of the kind $\langle \psi(\boldsymbol \theta) | \hat{h}_j \hat{g}_k | \partial_i \psi(\boldsymbol \theta)\rangle$ with $\hat{h}_j$ being $j$--th Hamiltonian term (unitary SWAP operator) and $\hat{g}_k$ being $k$--th unitary permutation. Note that any $\hat{g}_k$ permutation can be written as product of note more than $N - 1$ pair SWAP operators. The quantum circuit used to measure such quantity is shown in Fig.~\ref{fig:circuit}. The real and imaginary parts of $\langle \psi(\boldsymbol \theta) | \hat{h}_j \hat{g}_k | \partial_i \psi(\boldsymbol \theta)\rangle$ are measured using the Hadamard test protocol.

\begin{figure}[t!]
    \centering
    \includegraphics[width=0.49\textwidth]{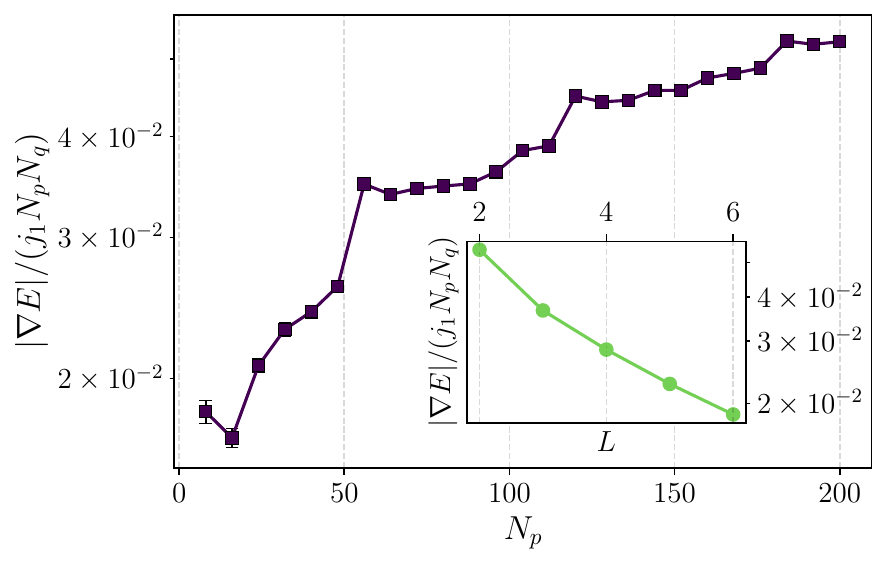}
    \caption{The gradient magnitude per parameter and qubit  $||\nabla E|| / (j_1 N_p N_q)$ within the $4 \times 4$ setup on the square lattice with $j_2 / j_1 = 0.4$ as a function $N_p$. Inset: $L \times 4$ setup on the square lattice at $j_2 / j_1 = 0.4$ as a function of $L$.}
    \label{fig:barren_plateaus}
\end{figure}

\subsection{Absence of barren plateaus}
\label{appendix:absence_of_barren_plateaus}
We investigate the possible presence or absence of barren plateaus in case of our Ansatz applied to Heisenberg models on two-dimensional frustrated spin-$1/2$ magnets. To this end, in Fig.\,\ref{fig:barren_plateaus} we plot the gradient magnitude per parameter and qubit $||\nabla E|| / (j_1 N_p N_q)$ (i) within the $4 \times 4$ square lattice setup at $j_2 / j_1 = 0.4$ as a function of $N_p$ and (ii) within the $L \times 4$ setup on square lattice at $j_2 / j_1 = 0.4$ as a function of $L$. To obtain the gradient magnitude, we individually draw each angle from the uniform distribution $\theta_k \sim \mathcal{U}(-0.0, 0.1)$, while the starting state is fully dimerized as suggested in Section\,\ref{sec:model_circuits}. We then average over a hundred of such random starting points. 

We observe that in the first setup, the average gradient per parameter increases and saturates as a function of the number of circuit parameters $N_p$. This is because a deep circuit allows one to rapidly change the outcoming quantum state by only small variations of the circuit parameters. In the meantime, in the second setup, the gradients decay sub-exponentially with the number of qubits $N_q$. Such behaviour was also reported on the case of the transverse field Ising model~\cite{PRXQuantum.1.020319}. Also, our observation is in line with the circuit architectures applied to frustrated two-dimensional spin-$1/2$ Heisenberg magnets~\cite{https://doi.org/10.48550/arxiv.2108.08086,Liu_2019}.

\begin{figure}[b!]
    \centering
    \includegraphics[width=0.49\textwidth]{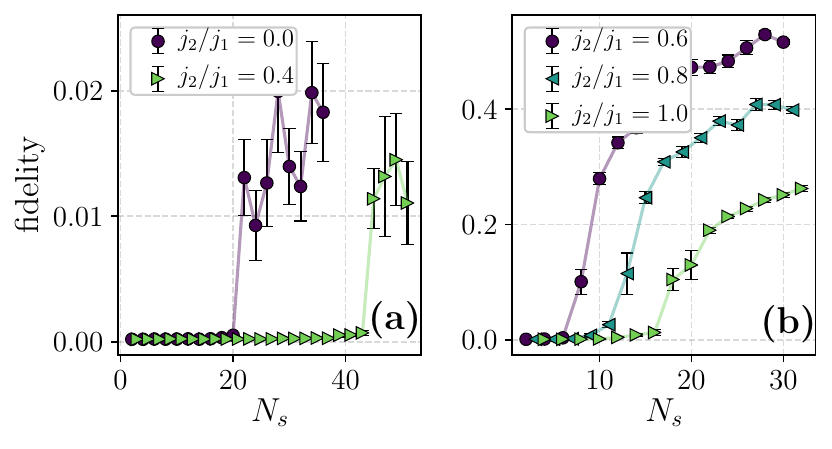}
    \caption{Fidelity as a function of $N_s$ for various $j_2 / j_1$ measured on the {\bf (a)} $2 \times 3$ kagome lattice (18 sites) with PBC and on the {\bf (b)} $4 \times 4$ triangular lattice with PBC.}
    \label{fig:appendix_transitions}
\end{figure}

\subsection{Algorithmic phase transition details}
\label{appendix:distribution_overlaps}
\begin{figure}[t!]
    \centering
    \includegraphics[width=0.49\textwidth]{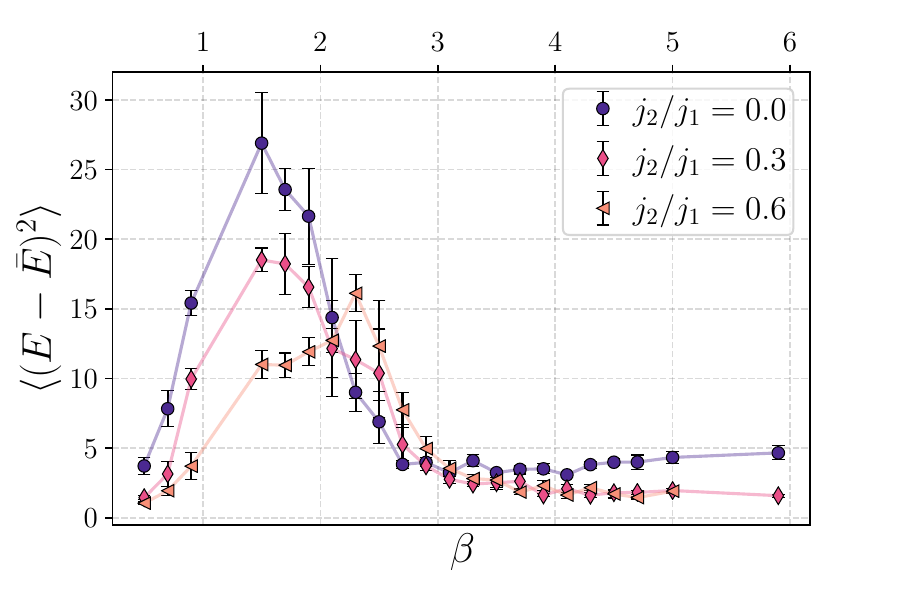}
    \caption{Energy variance $\langle (E - \hat{E})^2 \rangle$ as a function of $\beta$ on the $4 \times 4$ square lattice within sampling from the classical parametrical partition function $\mathcal{Z}_{\boldsymbol \theta}.$ The positions of maxima correspond to transition between the zero-overlap and trainable regimes.}
    \label{fig:cv_therm}
\end{figure}

In this Appendix, we provide additional details concerning observed algorithmic phase transition. In Fig.\,\ref{fig:appendix_transitions} we plot fidelity as a function of $N_s$ in two additional frustrated lattices: kagome and triangular. We observe similar clear separations of the regions with zero and non-zero fidelity. In Fig.\,\ref{fig:cv_therm} we plot energy variance $\langle (E - \hat{E})^2 \rangle$ obtained on the $4 \times 4$ square lattice within sampling from the classical parametrical partition function $\mathcal{Z}_{\boldsymbol \theta}.$ The maxima of the energy variance coincide precisely with the separation between zero-overlap and and trainable phases observed in Fig\,\ref{fig:panel1}\,(c).

\begin{figure}[b!]
    \centering
    \includegraphics[width=0.49\textwidth]{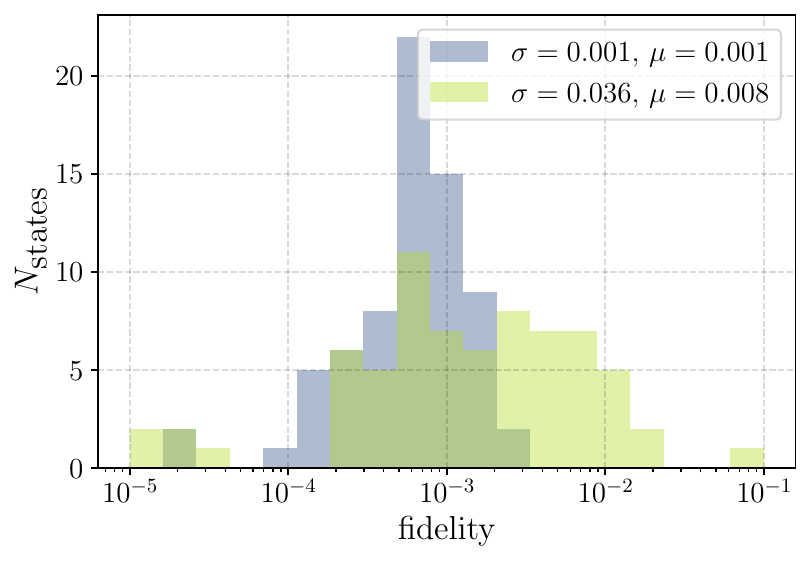}
    \caption{Histogram of overlaps with excited states $O_k = |\langle \psi_k | \psi(\boldsymbol \theta) \rangle|^2$ for $k$ corresponding to the lowest 100 states in the $S = 0$ sector measured at $j_2 / j_1 = 0.4$ on $4 \times 4$ square lattice with PBC at $N_s = 6$ and $N_s = 8$.}
    \label{fig:appendix_histogram}
\end{figure}

To get further insight into the algorithmic transition, in Fig.\,\ref{fig:appendix_histogram} we show histograms of fidelity of the first hundred $S = 0$ low-lying excited states above and below transition ($N_s = 6$ and $N_s = 8$) at $j_2 / j_1 = 0.4$ on the $4 \times 4$ square lattice. We see that below the transition, the histogram is very narrow with the variance $\sigma = 10^{-3}$ and the small average overlap of $\mu = 10^{-3} \approx 1 / |\mathcal{H}|$ with $|\mathcal{H}|$ being the $S = 0$ Hilbert space size of the problem. This indicates that all states that can possibly have overlap with our Ansatz wave function $|\psi(\boldsymbol \theta)\rangle$ are nearly of the same fidelity with a small variation. On contrary, after the algorithmic transition, the histogram widens to $\sigma = 0.036$, showing {\it selectivity} and that only few low-energy states dominate fidelity. This dramatic change of the histogram for $N_s = 6$ and $N_s = 8$ highlights a significant change in the stochastic VQE optimization performance upon passing the algorithmic phase transition.

\subsection{Number of iterations to saturation}
\label{appendix:N_SGD}
\begin{figure}[t!]
    \centering
    \includegraphics[width=0.49\textwidth]{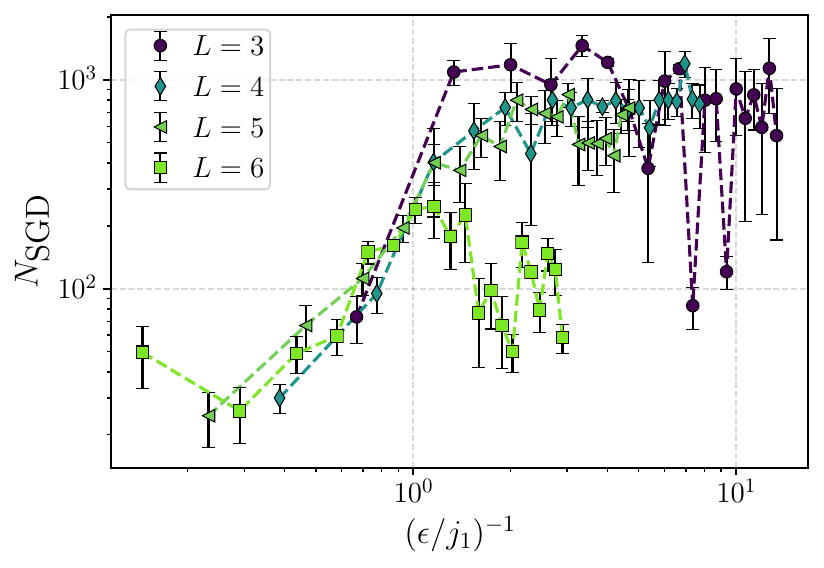}
    \caption{The number of SGD steps required to reach 90\,\% of the final fidelity $F_{\infty}$ within the $L \times 4$ setup on the square lattice at $j_2 / j_1 = 0.4$ as a function of inverse energy fluctuation $(\epsilon / j_1)^{-1}$.}
    \label{fig:N_SGD}
\end{figure}

In Fig.\,\ref{fig:N_SGD}, we investigate the number of SGD optimization steps required to reach 90\,\% of the final fidelity $F_{\infty}$ within the $L \times 4$ setup on the square lattice at $j_2 / j_1 = 0.4$ as a function of inverse energy fluctuation characteristic. We observe no pronounced dependence of $N_{\text{SGD}}(N_s)$ as a function of $L$ at a given $\epsilon$.

This contributes to our argument that the total required amount of computational resources $N_{\text{total}}(N_s) = N_s \times N_{\text{SGD}}(N_s)$ required to train a VQE wave function above the algorithmic phase transition, scales only polynomially with the system volume and thus the observed ``noisy'' phase with near-zero overlap can be avoided on larger clusters with an affordable circuit shots budget.

\begin{figure}[b!]
    \centering
    \includegraphics[width=0.49\textwidth]{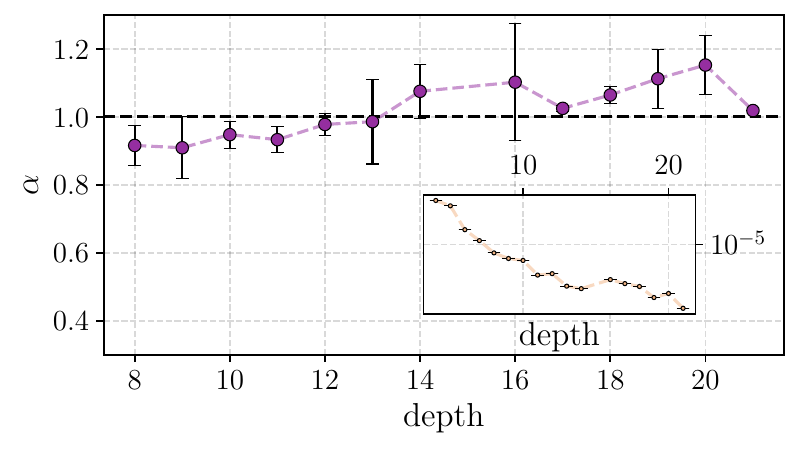}
    \caption{Depth dependence of the power $\alpha$ at $j_2 / j_1 = 0.4$ on a $4 \times 4$ square lattice. (Inset): $\mathcal{I}_0$ as the function of depth.}
    \label{fig:I0_alpha}
\end{figure}

\subsection{Large--$N_s$ fit parameters}
\label{appendix:fit_parameters}
In the main text, we fit the overlap with the empirical Ansatz $\mathcal{I}(N_s) = C / N_s^{\alpha} + \mathcal{I}_0.$ The $\alpha$ power was introduced to verify the proposed $1 / N_s$ scaling law. In Fig.\,\ref{fig:I0_alpha} we show, $\alpha$ and $I_0$ (inset) as functions of $D = N_p / 8$ for $j_2 / j_1 = 0.4$. Inability to express the ground state $I_0$, expectedly, decays exponentially with $D$. In the meantime, $\alpha$ remains close to $1$ within errorbars.

\subsection{Overlaps distribution in the large--$N_s$ setup}
\label{appendix:overlaps_distribution}

In this appendix section, in order to study the approximate $1 / \Delta^2$ scaling of the infidelity prefactor, we aim at distribution of the overlap over all excited states in the system. 

\begin{figure}[h!]
    \centering
    \includegraphics[width=0.49\textwidth]{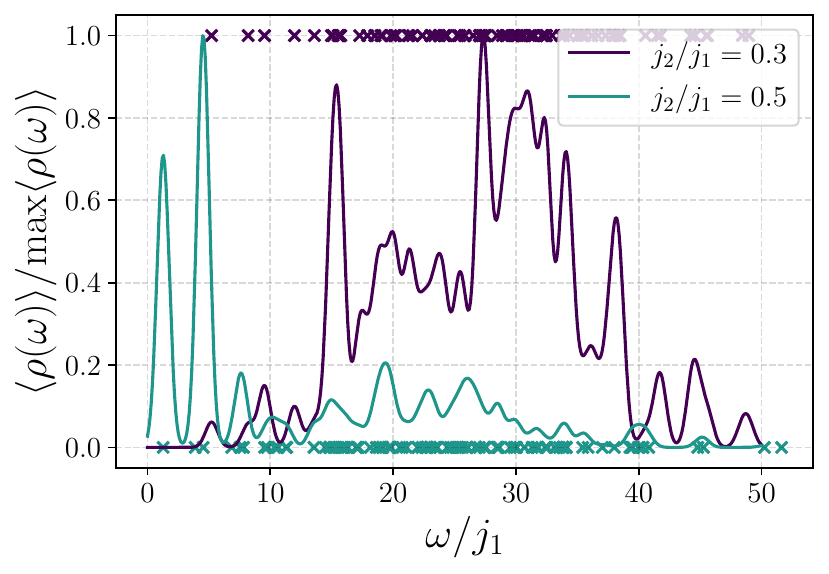}
    \caption{The normalized spectral function $\rho(\omega) = \sum\limits_k \delta(\omega - \Delta E_k) |\langle \psi(\boldsymbol \theta) | \psi_k \rangle|^2$ for $j_2 / j_1 = 0.3,\,0.5$ on the $4 \times 4$ square lattice. The delta-functions are smeared using the Gaussian approximation with the width $\Gamma / j_1 = 0.5$. For $j_2 / j_1 = 0.3$, on the $y = 1$ horizontal axis we mark the excited states energies, for $j_2 / j_1 = 0.5$ we do similarly at $y = 0.$}
    \label{fig:appendix_overlaps_spectrum}
\end{figure}

Considering translation symmetry-projected $4 \times 4$ square lattice, in Fig.\,\ref{fig:appendix_overlaps_spectrum} we plot the normalized spectral function $\rho(\omega) = \sum\limits_k \delta(\omega - \Delta E_k) |\langle \psi(\boldsymbol \theta) | \psi_k \rangle|^2,$ where $|\psi_k\rangle$ is the $k$--th excited state of the system, $\Delta E_k$ is its energy distance to the ground state, $|\psi(\boldsymbol \theta) \rangle$ is the variational state and the overlap is averaged over numerous SR optimization steps after convergence. The delta-functions are smeared using the Gaussian approximation with the width $\Gamma / j_1 = 0.5$. For $j_2 / j_1 = 0.3$, on the $y = 1$ horizontal axis we mark the excited states energies, for $j_2 / j_1 = 0.5$ we do similarly at $y = 0$. 

In the $j_2 / j_1 = 0.3$ case, the infidelity density is distributed relatively evenly over the excited states continuum in the middle of the Hamiltonian spectrum with energies much higher than the one of the first excited state. 
Recall that in the square lattice case at $j_2 / j_1 \to 0.6$, the system gap nearly closes, leading to the growing total infidelity of the variational state. We see that at $j_2 / j_1 = 0.5,$ the system gap shrinks significantly as compared to the $j_2 / j_1 = 0.3$ case and the dominant infidelity contribution is distributed only over several low-lying excitations, and not over the states continuum. Thus, as the system gap shrinks, the overlap with the first excited state and the other low-lying excited states grows. We also observe the growing density of the low-lying excited states as $j_2 / j_1 \to 0.6$. This interplay of the two factors (shrinking system gap and the growing low-energy excited states density) in the two-dimensional frustrated magnets might lead to the observed approximate $1 / \Delta^2$ growth of resources demand.

\bibliography{refs}
\end{document}